\documentclass{bioinfo}
\copyrightyear{}
\pubyear{}
\vol{}

\usepackage{graphicx}
\usepackage{epsfig}
\usepackage[]{url}
\usepackage{natbib}
\usepackage{algpseudocode}
\usepackage{subfigure}
\usepackage{graphicx}
\usepackage{hyperref}
\usepackage{color}
\usepackage{verbatim}
\usepackage{graphicx}
\usepackage{amsmath}
\usepackage{amssymb}
\usepackage{algorithm}
\newtheorem{theorem}{Theorem}

\newtheorem{definition}{Definition}

\newcommand{\bitem}{\begin{itemize}}
\newcommand{\eitem}{\end{itemize}}
\newcommand{\benum}{\begin{enumerate}}
\newcommand{\eenum}{\end{enumerate}}

\newcommand{\1}{{\bf 1}}

\newcommand{\bea}{\begin{eqnarray}}
\newcommand{\eea}{\end{eqnarray}}
\newcommand{\beas}{\begin{eqnarray*}}
\newcommand{\eeas}{\end{eqnarray*}}
\newcommand{\txtemail}[1]{\textbf{\texttt{#1}}}
\begin{document}
\firstpage{1}
\title[Lossy Compression of Quality Values via Rate Distortion Theory]{Lossy Compression of Quality Values via Rate
Distortion Theory}
\author[Asnani \textit{et~al}]{Himanshu Asnani$^{1}$, Dinesh Bharadia$^{1}$, Mainak Chowdhury$^{1}$, Idoia Ochoa$^{1}$, Itai Sharon$^{2}$ and Tsachy Weissman$^{1}$}
\address{$^{1}$Department of Electrical Engineering, Stanford University, Stanford, CA-94305, USA \\
$^{2}$Department of Earth and Planetary Science, University of California, Berkeley, CA-94720, USA}
\history{Received on XXXXX; revised on XXXXX; accepted on XXXXX}
\editor{Associate Editor: XXXXXXX}
\maketitle
\begin{abstract}
\section{Motivation: } Next Generation Sequencing technologies
revolutionized many fields in biology by
enabling the fast and cheap sequencing of large amounts of genomic
data. The
ever increasing sequencing capacities enabled by current sequencing
machines
hold a lot of promise as for the future applications of these
technologies,
but also create increasing computational challenges related to the
analysis
and storage of these data. A typical sequencing data file may occupy
tens or
even hundreds of gigabytes of disk space, prohibitively large for
many users.
Raw sequencing data consists of both the DNA sequences (reads) and
per-base
quality values that indicate the level of confidence in the readout
of these
sequences. Quality values account for about half of the required disk
space
in the commonly used FASTQ format and therefore their compression can
significantly reduce storage requirements and speed up analysis and
transmission of these data.
\section{Results: }
In this paper we present a framework for the lossy compression of the
quality value sequences of genomic read files. Numerical experiments
with reference based alignment using these quality values suggest that
we can achieve significant compression with little compromise in
performance for several downstream applications of interest, as is
consistent with our theoretical analysis. Our framework also allows
compression in a regime - below one bit per quality value -
for which there are no existing compressors.
\section{Availability:\url{http://www.stanford.edu/~mainakch/svdbit.tgz}}
\section{Contact: \txtemail{asnani@stanford.edu,dineshb@stanford.edu,\\
mainakch@stanford.edu,iochoa@stanford.edu,\\
itaish@berkeley.edu,tsachy@stanford.edu}}
\end{abstract}

\section{Introduction}
\label{sec::introduction}
\subsection{Background and Motivation}
\label{subsec::motivation}

It has been more than a decade now since the first draft of the human
genome
was published \citep{Lander2001}. The Human Genome Project, which
required a
significant collaborative effort of many scientists for more than
$10$ years,
was completed using the Sanger sequencing technology and is estimated
to have
cost almost three billion dollars. Just a decade later, many medium
and small
size laboratories achieve the task of sequencing complete mammalian
genomes
within a few weeks using the new next generation sequencing (NGS)
technologies.
Read size is usually smaller for NGS sequencers compared to Sanger
sequencing,
but sequencing throughput is significantly higher. Current sequencers are
capable of generating close to tera-base worth of data that needs to
be stored
and processed. Several recent studies, such as the cow rumen
\citep{Hess2011} and the MetaHit \citep{Qin2010} metagenomic
projects resulted with
hundreds of hundreds of giga-base worth datasets. As project scales will
continue to grow, it is expected that the bottleneck of projects involving
massive sequencing will move towards the computational aspects, in
particular with respect to the analysis and storage of the data. As
a result,
there is a growing interest in computational tools that can speed up
processing and compressing this type of data.
\begin{figure}[htbp]
\begin{flushleft}
\includegraphics[width=\columnwidth]{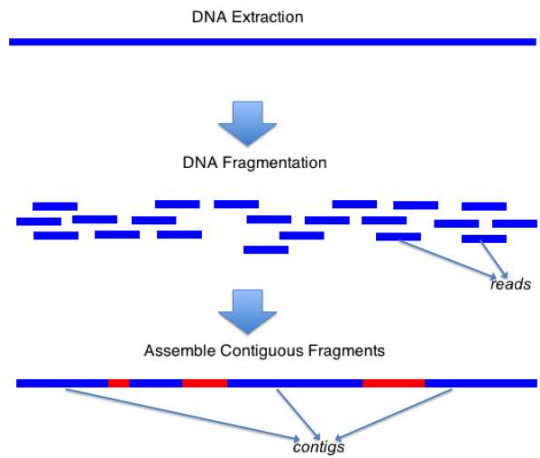}
\caption{Essential rubrics of gene-sequencing.}
\label{fig::gene-sequencing}
\end{flushleft}
\end{figure}
\par
The sequencing process begins with the shearing of the input DNA into
many short pieces, which are then prepared for sequencing, loaded onto the
sequencer and sequenced. (Figure \ref{fig::gene-sequencing}). Different
methods are used by the different
NGS technologies for the readout of the sequencing signal (also known
as base
calling). This process may be interfered by various factors, which may
lead to
wrong readout of the sequencing signal. In order to assess the probability
for
base calling mistakes, sequencers produce scores that reflect the level of
confidence in the readout of each base. These scores are known as quality
values and are part of the standard sequencing output. In the widely
accepted
FASTQ format, for example, each read is represented by four lines,
of which
two are for the reads themselves and a string of quality values for
the read.
The use
of quality values depends on the use of the data. Low quality reads
and read
parts may be removed from the data prior to operations that require high-
quality data such as the assembly of genomes, or mapping-based single
nucleotide polymorphism (SNP) detection.
Next we describe in detail the quality value sequences and the FASTQ format.
\subsection{FASTQ format}
\label{subsection:fastq}

We consider the compression of FASTQ files, because of their
wide acceptance as a standard for storing sequencing data. A
FASTQ file consists of separate entries for each read, each one
consisting of
the following four lines (see Figure \ref{tab1} for example):

(i) header line, always begins with an “@” sign, followed by the
name of the read

(ii) $r$ - the base pair sequence in the read, where $r[i]\in\{A,C,G,T,N\} $
with $N$ representing an unknown base.

(iii) quality value header, begins with a “+” sign which may or may
not be
followed by the read name

(iv) $q$ - the quality value string for the sequence $r$. $q[i]$ represents the
quality value of base $r[i]$.

\begin{figure}
\begin{align*}
	&\text{\texttt{@SRR001666.1}}\\
	&\text{\texttt{GATTTGGGGTTCAAAGCAGTGCAAGC}}\\
&+\\
&\text{\texttt{!''*((((***+))!!+++!!)(!(( } }
\end{align*}
\caption{Typical FASTQ record. We focus on lossily compressing the
fourth line
- the quality value string - of every record.}
\label{tab1}
\end{figure}
\par
Quality value $q[i]$ represent the level of confidence in the readout of its
corresponding base $r[i]$, with high quality values representing greater
confidence. Each quality value is encoded by an ASCII character in the
FASTQ format based on one of a few accepted schemes.
One such standard is the Sanger scale
\citep{Cock2009}. Quality values in
this scale are computed as $Q = 33-10 \log_{10} P$, and range typically
from
$33-73$.
Lossless compression of read files will require, on average, $2$
bits/symbol
for the base sequence, and $6$ bits/symbol for the quality values,
i.e., three
times as much storage as is required for the reads themselves. Base
calling
is an inherently noisy process by itself. Based on the amount of noise added by
this process, it might be possible to achieve a significant reduction in the
representation of these values with only a marginal loss of performance by
neglecting a the portion of the quality value information that encodes mostly
noise.  Here we explore the use of lossy compression for achieving this goal.

\subsection{Related Work and Our Contribution}
\label{subsec::contribution}
The literature abounds in efforts to compress the genomic data. Several
approaches exist for compression of whole genomes without the aid of any
external information, see for example \citep{Chen1999}, \citep{Chen2002},
\citep{Cao2007}, \citep{Sato2001}, \citep{Pinho2011} and references
therein.

Whole genome level compression without the aid of any external information has
been the focus of \citep{Chen1999} - \citep{Pinho2011} and references
therein.
More recent contributions show that further compression can be achieved by
mapping the target genome to a reference genome and encoding only the
differences between the two \citep{Christley2008},\citep{Pinho2011a},
\citep{Wang2011},
\citep{Kuruppu2010}, \citep{Kuruppu2011}, \citep{Heath2010},
\citep{Ma2012}, \citep{Chern2012}.
Other approaches consider the problem of losslessly compressing the
sequence reads
together with the corresponding quality values \citep{Deorowicz2011},
\citep{Tembe2010}, while in \citep{Timothy2008} only the compression of
the reads is
considered.
\par
In this paper we concentrate only on the lossy compression of the
quality values
as they take up chunk of the storage space.  Lossy compression of
quality values
have been considered in the literature only recently. It has been
presented as a
plausible idea in \citep{Leinonen2011} and a concrete algorithm as a
part of SLIMGENE Package in \citep{Kozanitis2011}, which
considered the problem of compression of the reads, both of the base
sequences
and quality values sequences. Their compression package has a module
which does
a lossy compression of quality values, based on fitting a fixed state
markov
encoding model on adjacent gaps between quality scores. They use SNP
variant
calling as their performance metric and show lossy compression has
a minimal
effect on performance. In  \citep{Fritz2011}, a metric called
``quality-budget" is used to selectively discard the quality values
which match
perfectly to the reference, with only quality values with sufficient
variation
being retained. Recently,  \citep{Wan2011}, considered relative
mapping accuracy of the reads as the metric and applied various lossy
transform
techniques to show that impressive compression can be achieved with
similar
mapping between uncompressed and compressed quality value files. The importance
of lossy compression schemes in the context of the differential archiving needs
of sequenced data
has been highlighted in \citep{Cochrane2012}.
\par
In this work, we use Rate Distortion theory to guide our construction
of lossy compressors. While details are deferred to later sections, in
general, Rate Distortion theory pertains to compressing or representing
an information source with a certain number of bits per source component
(rate)	while minimizing the distortion between the source and the
reconstruction based on said representation, as measured by a given
fidelity criterion.  As
already alluded to, due to their inherently noisy nature, some of the
quality values can
be discarded or represented in a coarser resolution without incurring much
performance loss for downstream applications. That is, that an
appropriately quantized file
should give a performance at a downstream application comparable to the
unquantized or original file. The distortion between the uncompressed
(i.e., original quality values)  and
compressed source (i.e., the reconstructed quality values after the
lossy compression) data is a mathematical
quantity such as hamming loss, square loss, etc, rather than ``physical
distortion'' or performance loss due to lossy compression with
respect to downstream applications such as Alignment, SNP Calls, etc. Hence, the question arises, can we compress the quality value file
to a
certain rate (read, bits, etc.) without incurring much loss with respect
to a standard distortion criterion and hope that a low distortion would
imply little
compromise in performance at downstream applications? In other words,
does our
fidelity criterion correspond well to ``physical
distortion''?
\par
Towards answering this question, we use mean square error as the
distortion
criterion for our lossy compression.  We choose to work with this
particular distortion criterion due more to its convenient analytical
properties than to our belief that it is canonical to measuring the loss
incurred in the downstream applications.  Further, we  model our source
of the quality values in the read file as a multivariate gaussian. This
is justified by  both central limit arguments (as the sources of noise
in the acquisition of the quality values are incremental and independent)
and  the fact that,
given a vector source with a particular co-variance structure, the
Gaussian
Multivariate source is the least compressible  and, further, a code
designed under the Gaussian assumption will perform at least as well on
any other source of the same covariance \citep{Lapidoth1995}.  We then
suggest a tractable scalar quantization algorithm and
show numerically that it achieves mean square loss comparable to
the optimum (that would be achieved using the optimal vector
quantization). We further demonstrate that achieving low mean square
error translates to comparable performance in the downstream applications
as compared to
use of the original (uncompressed) quality values.

Our algorithm operates at any non-zero compression rate, and as far as
we know  is the
first implementation of lossy compression of quality values that can
accommodate less than one bit per quality value. We find reasonable
performance in the downstream applications even in this low-rate regime.

\subsection{Organization of the Paper}
\label{subsec::organization}
The paper is organized as follows. In Section \ref{sec::problem}, a
problem formulation is provided with emphasis on
how the quality values are modeled, the class of schemes which are
considered, and the
performance metrics used. Section
\ref{sec::rate-distortion} provides some background
on Rate Distortion theory for memoryless sources. Our primary compression
technique is transform coding via
\textit{singular value decomposition} (SVD), which we describe in
Section \ref{sec::transform-coding}. Experiments
on real data are presented in Section \ref{sec::results}. The paper is
concluded in Section in \ref{discussions} with directions for future work.

\section{Problem Formulation}
\label{sec::problem}
We now formalize the problem of lossy compression of
quality values and describe the general model.
As discussed in Section \ref{subsec::motivation}, quality values
represent the
reliability of a particular read base. The higher
the quality value, the higher the reliability of the corresponding
read base,
and vice versa. More specifically quality value is the integer mapping
of $P$ (the probability that the corresponding
base call is incorrect) and is represented in (at least) the following
different
scales/standards :
\begin{itemize}
\item \textit{Sanger or Phred} scale : $Q=-10\log_{10}P$.
\item \textit{Solexa} scale : $Q=-10\log_{10}\frac{P}{1-P}$.
\end{itemize}
The integer $Q$ values are encoded in ASCII format, for the purpose of
this work, and without loss of generality, we consider the  \textit{Phred}$+33$
in which the range of quality values is $[0,40]$. A quality value $QV$ is
represented by the letter whose ASCII value is $QV+33$, resulting with letters
in the ASCII range of $[33, 73]$.

\subsection{Modeling Quality Values}
\label{subsec::modeling}
We consider files with fixed read length, $l$. Denote the
number of
reads in the file by $N$. The quality values
in a file are denoted by $\{\mathbf{X}_i\}_{i=1}^{N}$, where
$\mathbf{X}_i=[X_i(1),\cdots,X_i(l)]$.	In real data quality values
take integer values in a finite alphabet, $\mathcal{X}$, for example in
\textit{Phred}$+33$ scale, $\mathcal{X}=\{33,34,\cdots,72,73\}$.
However, for the purpose of modeling, we assume $\mathcal{X}=\mathbf{R}$
(the set
of real numbers). Each $\mathbf{X}_i$
is modeled as independent and identically distributed jointly Gaussian
random
vector distributed  as
$\mathcal{N}(\mu_{\mathbf{X}},\Sigma_{\mathbf{X}})$.
 The motivation for modeling the reads as Gaussian is already outlined
 in Section \ref{subsec::contribution}, while independence assumption
 is supported by the fact that reads, in general, are randomly sampled
 from the genome in gene-sequencing step.

\subsection{Scalar Quantization}
\label{subsec::scalar-quantization}
\begin{figure}[htbp]
\begin{center}
\includegraphics[width=\columnwidth]{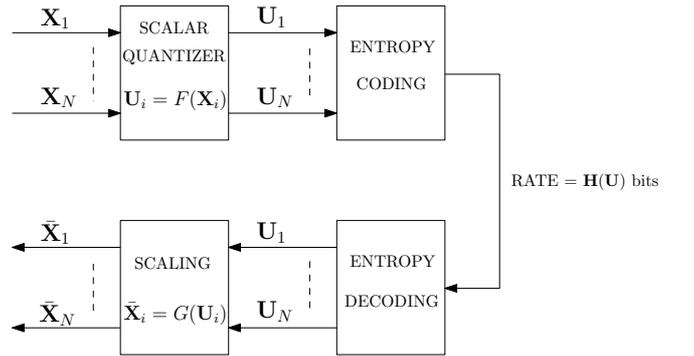}
\caption{Lossy Compression of Quality Values via scalar quantization.}
\label{scalar-quantization}
\end{center}
\end{figure}

The compression techniques which are applied and analyzed in this paper,
can be
modeled as in Fig. \ref{scalar-quantization}.

The quality value vectors $\mathbf{X}_i$ are i.i.d. as
$P_{\mathbf{X}}\sim\mathcal{N}(\mu_{\mathbf{X}},\Sigma_{\mathbf{X}})$,
and each is  mapped
into decision regions representable
in a finite number of bits. This is the quantization step represented by
$F(\mathbf{X}_i)$. Let the
mapped (quantized) values so obtained be referred to by $\mathbf{U}_i$.
$\mathbf{U}_1^N = (\mathbf{U}_1, \ldots , \mathbf{U}_N)$ are then
losslessly  described via
a lossless encoder (such as a Huffman encoder, LZ encoder, etc.). The
rate $R$ of
compression is the average number of bits per
quality value used to describe the reads. This completes the
compression step and is
lossy in general due to the quantization step. The quantization could be as simple as truncating some entries, or
rounding, or more sophisticated, such as transform coding which is the
focus of this paper and
will be described in Section \ref{sec::transform-coding}. We refer to
this as ``scalar quantization'' because each read is quantized
separately, unlike vector quantization techniques where different reads
are collected and jointly quantized. Vector
quantization  generalizes scalar quantization, but is harder to implement
and typically has minor performance improvements.
\par
For reconstructing the quality value sequence, entropy decoding is used to
losslessly reconstruct $\mathbf{U}_1^N$
from its bit description, and then the mapping $G(\cdot)$ is used to
get a lossy estimate of $\mathbf{X}_i$ as
$\overline{\mathbf{X}}_i$.
The normalized distortion satisfies
\bea
D=\frac{1}{N}\sum_{i=1}^N\frac{1}{l}\parallel
\mathbf{X}_i-\overline{\mathbf{X}}_i\parallel_2^2\stackrel{N\rightarrow\infty}{
\longrightarrow}\frac{1}{l}\mathbf{E} \left[\parallel
\mathbf{X}-\overline{\mathbf{X}}\parallel_2^2\right],
\eea
where the limiting behavior as $N\rightarrow\infty$ is due to the law
of large
numbers and $\mathbf{E}$ is
the expectation over the statistics of source $\mathbf{X}$. There is
obviously a
tradeoff between rate and distortion, which depends on the source
$\mathbf{X}$ and is quantified via the scalar rate-distortion function,
$R^{(s)}(D,\mathbf{X})$ or distortion-rate function
$D^{(s)}(R,\mathbf{X})$ (the superscript s indicates scalar quantization).
Henceforth, the compression model in the paper is
that of Fig. \ref{scalar-quantization}, which is a special case of the
general compression architecture that Rate Distortion theory allows,
as is briefly reviewed in Section
\ref{sec::rate-distortion}. We demonstrate empirically in Section
\ref{sec::transform-coding} that scalar quantizers for our data work
as well as any vector quantizer. Also, from now on the terms
`quantization'
and `compression' are used interchangeably.

\subsection{Performance Metrics}
\label{subsec::performance}
As discussed in Section \ref{subsec::contribution}, the
mean square distortion is mathematically convenient. Of interest in
practice is the deterioration in performance when using the lossily
reconstructed quality values  relative to a
file containing the original quality values.
A genome sequence read file, i.e., the  FASTQ file can be utilized in a
variety of
applications by a number of downstream
analyzers in genomics. However, almost all such downstream applications
would
eventually depend on the alignment profile of the reads. This alignment
may be
either \emph{de novo} or \emph {reference based}. We consider reference
based
alignment here and describe two natural performance
metrics that will be used for comparison of our scheme with
other schemes. They are :

\begin{enumerate}
\item \textit{Relative Mapping Accuracy}: Let us denote a typical
base sequence
present in the fastq file by $r$.
The position where $r$ maps to the reference  is denoted by
$\mathcal{P}(r)$.
Let $\mathcal{A}=\{(r,\mathcal{P}(r)):r\in\mbox{\ FASTQ}(\mathbf{X})\}$
or the file containing the original uncompressed quality values. From
now on
with some benign abuse of notation
we will abbreviate the original fastq file with uncompressed quality
values
simply as the uncompressed fastq file.
Similarly we denote
$\overline{\mathcal{A}}=\{({r},{\mathcal{P}}({r})):{r}\in{\mbox{\
FASTQ}}(\overline{\mathbf{X}})\}$
for the compressed fastq file. The relative mapping accuracy is simply
\begin{equation}
\frac{|\mathcal{A}\cap \overline{\mathcal{A}}|}{|\mathcal{A}\cap
\overline{\mathcal{A}}|+|\mathcal{A}\setminus \overline{\mathcal{A}}|}
\label{eq:relmap}
\end{equation}
where $\cap$ stands for intersection and $\mathcal{A}\setminus \mathcal{B}$ denotes the elements that belong to $\mathcal{A}$ but not to $\mathcal{B}$.

In other words, the relative mapping accuracy measures what percentage
of original (uncompressed) reads have mapped to the same position on
the reference sequence with the lossily compressed quality values.

\item \textit{Symmetric Difference}: This is simply
\begin{equation}
\frac{|\mathcal{A}\setminus
\overline{\mathcal{A}}|+|\overline{\mathcal{A} }
		\setminus \mathcal{A}|}{|\mathcal{A}\cap
\overline{\mathcal{A}}|+|\mathcal{A}\setminus \overline{\mathcal{A}}|}
\label{eq:symmdiff}
\end{equation}
This measures the percentage of reads that align to different positions
on the reference sequence with the uncompressed and the lossily compressed
quality values.
\end{enumerate}
Ideally, we would want our read file containing compressed quality values
to give an alignment profile identical to that of the
read file with the uncompressed quality values. In other words, we want a high relative mapping
accuracy and a low symmetric difference.

\section{Rate Distortion Theory : Some Preliminaries}
\label{sec::rate-distortion}
\begin{figure}[htbp]
\begin{center}
\includegraphics[width=\columnwidth]{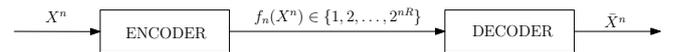}
\caption{Rate-distortion problem}
\label{fig::rate-distortion}
\end{center}
\end{figure}

In this section, we provide a brief background on Rate Distortion
theory for
memoryless sources.  For detailed description and proofs please refer to
\citep{Cover1991}.  We consider fixed rate schemes which are as
follows. Referring to Fig. \ref{fig::rate-distortion}, our goal is to
encode a
source sequence of block length $n$, $X^n$, using only $nR$ bits, in
order to
minimize the distortion between the original source sequence and the
reconstruction sequence, $\overline{X}^n$, chosen
by the decoder. We assume that our given distortion function $d : (X , \overline{X})
\rightarrow \mathcal{R}^+$ operates symbol by symbol (as opposed to
block by
block) and that the distortion $D$ is
given by $D = \mathbf{E}\left[d(X^n
, \overline{X}^n)\right] = \mathbf{E}\left[\frac{1}{n}\sum_{i=1}^n d(X_i,
\overline{X}_i)\right]$.
\begin{definition}
A rate-distortion scheme of rate R consists of the following:
\begin{enumerate}
\item An encoder, $f_n : X^n\rightarrow\{1, . . . , 2^{nR}\}$.
\item A decoder, $g_n : \{1, . . . , 2^{nR}\} \rightarrow \overline{X}^n$.
\item A reconstruction sequence, $\overline{X}^n = g_n(f_n(X^n))$.
\end{enumerate}
\end{definition}
\begin{definition}
The pair $(R,D)$ is said to be achievable if $\forall \epsilon > 0$,
$\exists\ n$
and a rate-distortion scheme at rate $\le R+\epsilon$ and (expected)
distortion
$\le  D + \epsilon$.
\end{definition}
\begin{definition}
The rate-distortion function is defined as $R(D,\mathbf{X})=\inf\{R':
(R', D)
\mbox{ is achievable}\}$. Similarly, we define the distortion-rate
function as
$D(R,\mathbf{X}) = \inf\{D': (R, D') \mbox{ is achievable}\}$.
\end{definition}
\begin{theorem}
\label{theorem2}
 Gaussian Memoryless Scalar Source \citep{Cover1991}: For an i.i.d.
Gaussian scalar source $X \sim $ $\mathcal{N}(\mu,\sigma)$, the
rate-distortion and the distortion-rate functions are:
\bea
R(D,X)&=&\frac{1}{2}\log\left(\frac{\sigma^2}{D}\right)\1_{\{D<\sigma^2\}}\\
D(R,X)&=&\sigma^2 2^{-2R},
\eea
where $\1_{\{A\}}$ is the indicator function that takes the value one when the event $A$ is true and zero otherwise.
\end{theorem}
\begin{theorem}
 Gaussian Memoryless Vector Source with independent components
\citep{Cover1991}: For an i.i.d. Gaussian vector source ($\mathbf{X}$),
$\mathcal{N}(\mu,\Sigma_{\mathbf{X}})$, with
$\Sigma_{X}=\mbox{diag}[\sigma_1^2,\cdots,\sigma_l^2]$ (i.e., independent
components), the optimal distortion-rate tradeoff is given as the
solution to
the following optimization problem:
\bea
D(R,X)&=& \min_{\rho=[\rho_1,\cdots,\rho_l]} \Sigma_{i=1}^l \sigma_i^2
2^{-2\rho_i} \\
&& \mbox{s.t. } \Sigma_{i=1}^l \rho_i \le R.
\eea
\end{theorem}
\section{Compression via Singular Value Decomposition (SVD)}
\label{sec::transform-coding}
\begin{figure}[htbp]
\begin{center}
\includegraphics[width=\columnwidth]{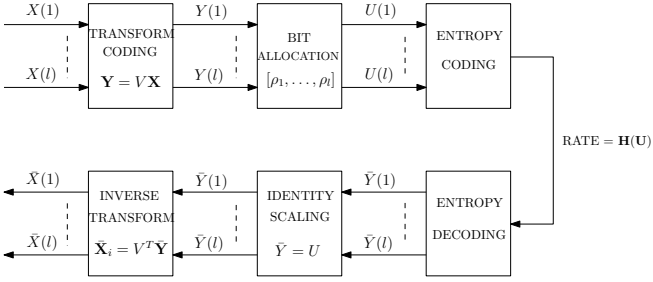}
\caption{The scalar quantization based on transform coding and
differential bit-allocation.}
\label{fig::transform-coding}
\end{center}
\end{figure}

The general transform coding paradigm is shown
in Fig. \ref{fig::transform-coding}.  If the source/signal is
``compressible''
in a particular domain, intuitively one should transform the source/signal
to
that domain.
First the read vector $\mathbf{X}$ (vector of length $l$), is decorrelated
by a
unitary operation matrix $V$ ($VV^T=I$) computed from the empirical
statistics
of $\mathbf{X}$. The Bit Allocation block then allocates bits to
each read
position. This allocation is precomputed  by using the statistics of
the quality
value file. Thus for each read, $\mathbf{Y}=V\mathbf{X}$ is quantized by a
scalar quantizer, to obtain bits $\mathbf{U}$ which are then finally
compressed
into bits by an entropy encoder.  To obtain the `quantized' read vector
$\overline{\mathbf{X}}$, the compressed bit description is first entropy
decoded and
then demapped by the quantizer into decisions for each bit sequence
followed by
inverse transform through $V^T$.
\par

\textbf{SVD-BitAllocate}: Here we perform transform coding as in Fig.
\ref{fig::transform-coding} with $V=V_{svd}^T$ .

The rate of bits allotted per quality value sequence is a user specified
parameter. The source would be compressed
accordingly.   Thus we can formulate the bit allocation problem as
a convex
optimization problem and solve it exactly. That is, given  a budget
of $R$ bits per read, we allocate the bits by first transforming
$\mathbf{X}$
into $\mathbf{Y}$ (a decorrelating transform, which by the Gaussian nature
makes the components independent) and then allocate bits to the
independent
components of $\mathbf{Y}$ by solving the following optimization problem,
\begin{align}
	\text{minimize}_{\rho} \frac{1}{l} {\sum_i \sigma^2
	2^{-\rho_i}}\notag \\
	\text { subject to } \sum_i\rho_i \leq R \label{eqopt}.
\end{align}
The objective function here is the mean squared error per quality value.
Hence the rate = $\sum_{i=1}^l\rho_i$, where $\rho_i $ is the number
of bits
allocated to the $i^{th}$ component of $\mathbf{Y}$. Since $\mathbf{Y}$
has
independent Gaussian components under our modeling assumption, the optimal
value of the above Problem \ref {eqopt} is exactly equal to the optimal
distortion rate function for a Gaussian source as outlined in Section
\ref{sec::rate-distortion}.  The solution to \ref{eqopt} dictates the
number of bits that needs to be allocated to store $Y_i$. Ideally this
allocation should be done by vector quantization for the whole block
$\emph{Y}_1^N$ together. However, due to ease of implementation and
negligible performance loss, we use a scalar quantizer. Thus the
component $Y_i$
is normalized to a unit variance Gaussian (the variances of each
component are
either known from the statistics of the read file or are estimated )
and then
it is mapped to decision regions representable in $\rho_i$ bits. The
decision
regions and their representative values (stored in
$\mathcal{D}_{map}(\rho_i)$
for all possible $\rho_i$)
are found from a Lloyd Max procedure on a scalar Gaussian distribution,
i.e.,
for $\rho_i$ bits the $\mathcal{D}_{map}(\rho_i)$ will store
$2^{\rho_i}$ regions (boundary points and representative value for
each region)
which would give the minimum mean squared error for a unit variance
Gaussian.

\begin{algorithm}
\caption{SVD-BitAllocate($X_1^N,R)$}
\begin{algorithmic}
\State $\mu_{\mathbf{X}},\Sigma_{\mathbf{X}} \gets $ Empirical mean and
covariance of $\mathbf{X}_1^N$
\State Compute SVD  $\Sigma_{\mathbf{X}}=V_{svd} S V_{svd}^T$,
$S=\text{diag}(\sigma_1^2,\ldots,\sigma_l^2)$
\State Precompute Lloyd Max quantizer $\mathcal{D}_{map}$ for gaussians
\For {$i = 1 \to N $}
\State $\mathbf{Y}_i \gets F_{svd}\mathbf{X}_i$
\State $ \rho^{\ast} \gets \mbox{\textit{BitAllocate}}(S,R)$
\State $\mathbf{U}_i\gets
\mbox{\textit{Scalar-Quantization}}(\mathcal{D}_{map},\mathbf{Y}_i,\rho)$
\EndFor\newline
\Function{BitAllocate}{$S,R$}
\State$\min_{\rho} \frac{1}{l}\sum_{i=1}^l \sigma_{i}^2 2^{-2\rho_i}$
 \State such that $\sum_{i=1}^l \rho_i\le R$
\EndFunction
\Function{Scalar-Quantization} {$\mathcal{D}_{map},\mathbf{Y}_i,S,\rho$}
\For {$j=1 \to l$}
\State$\tilde{Y}_i(j)\gets \frac{1}{\sigma_j}Y_i$
\State $\tilde{U}_i(j)\gets \mbox{ Quantize } \tilde{Y}_i(j) \mbox{
using }
\mathcal{D}_{map}(\rho_j)$
\State ${U}_i(j)\gets\sigma_j \tilde{U}_i(j)$
\EndFor
\EndFunction
\end{algorithmic}
\end{algorithm}
Once we get $\mathbf{U_1^N}$, we may perform lossless compression using
standard universal entropy coders. However, this was found to achieve
negligible
compression improvements over $R$ bits per read, and hence was not
considered in the numerical results, to which we now turn.
\section{Results}
\label{sec::results}
\begin{figure}
	\includegraphics[width=\columnwidth]{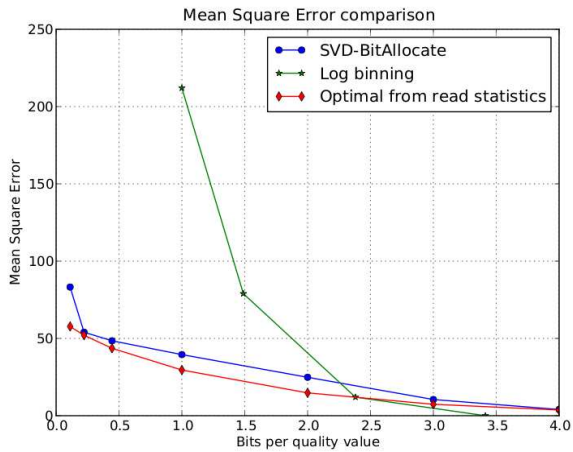}
	\caption{Mean squared error plots as a function of the number
	of bits
		allocated per position. The optimal here refers to
		the solution
		of Problem \ref{eqopt}. Log Binning is proposed in
		\citep{Wan2011}.}
	\label{fig:meansquare}
\end{figure}

We present the results of numerical experiments with our algorithm on
real read
data. The data was downloaded from the NCBI human genome sequence
read archive
\citep{NCBI2012} (reads with identifier ERR000531 are used). The total number
of quality value sequences considered for the data
presented is about 20 million, each sequence length (i.e., read length)
being 46.
We tested the mean squared error performance of the quantized
quality values from our algorithm against other algorithms \citep{Wan2011}
Figure
\ref{fig:meansquare}.
\begin{figure}
	\includegraphics[width=\columnwidth]{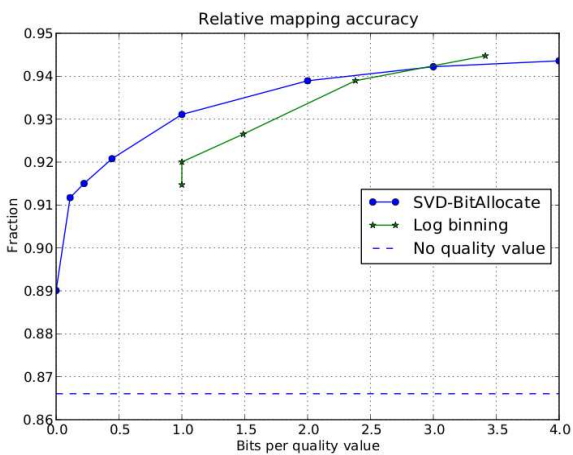}
	\caption{Percentage of reads which have been aligned to the same
		positions as the original unquantized reads. Log
		Binning has
been proposed in \citep{Wan2011}.}
	\label{fig:originalpercentage}
\end{figure}
\begin{figure}
	\includegraphics[width=\columnwidth]{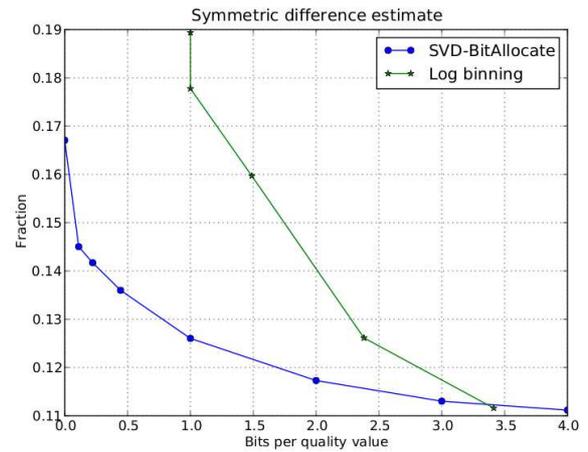}
	\caption{Fraction of size of symmetric difference over size
	of mapped
		reads with unquantized values. Log Binning has been
		proposed in
\citep{Wan2011}.}
	\label{fig:symmetricdifference}
\end{figure}
The results show that for low number of bits per quality value position,
our
algorithm achieves much lower mean squared error compared to existing
implementations (binning based quantizers \citep{Wan2011}). At higher
values, the
degradation in mean squared error (MSE) comes from the fact that our
modelling
assumption works with
continuous real numbers, hence the ``optimal'' mean square value does
not vanish
even with increasingly many bits. Thus, the log binning curve (which
works with the integers directly) performs better and actually goes
below the
``optimal'' MSE curve for sufficiently high number of bits. Note that
with 6
bits per position, we can code for the quality values losslessly.

We also show the alignment performance by a sequence read aligner
(bowtie)
with our quality value sequences. Figure \ref{fig:originalpercentage} and \ref{fig:symmetricdifference}
show the performance of the relative
mapping accuracy  and symmetric difference (defined in Equations
\ref{eq:symmdiff} and \ref{eq:relmap}) between reference
based alignment
using the quantized files as a function of the bits allocated per
quality value
position.
This can influence several downstream applications like variant calling
/SNP
detection (Figure \ref{fig:originalpercentage}). The plots show little
performance loss even with very small number of bits per position. This
also
corroborates our overall claim that lower distortion with respect to
mean square
loss translates to comparable performance in the downstream
applications. Further our
framework allows us to work with less than one bit per quality value,
which
may prove invaluable in future applications where the number of reads and
their lengths will be increased manyfold.
Also note that for zero rate, we are using just the mean of the
quality values over all the reads. Since this needs a constant
storage cost, the amortized number of bits required to store this
information for large numbers of reads is zero. The curves in Figure
\ref{fig:originalpercentage} suggest that even with this information,
we can achieve performances much better than by discarding quality
values altogether.

The plots, as expected, show increasingly better match with higher rates,
with reconstructions using the original quality value sequence. This is
due to the fact that the alignment performance has been compared to the
uncompressed values. However, in accordance with our conjecture to
be studied
in future work, the curve measuring the `true' performance with respect
to the
yet unknown `ground' truth may not be monotone with increasing rate, as
limiting the rate may denoise the data and hence enhance the accuracy.

\section{Discussion}
\label{discussions}
We have presented a scheme for lossy compression of the quality value
sequences
arising in  genomic data. By directly allocating bits to the most
significant variations, our scheme simply and effectively captures the
information in the quality value sequence given limited storage resources.
While refinements such as the use of clustering ideas to learn the
statistics of
the reads more finely would likely result in improved performance
in the downstream applications (as compared to using the original
quality
value sequence), we suspect, based on preliminary observations, that our
scheme may also achieve
some form of denoising. Our thesis is that appropriate lossy compression
of the quality values may result not only in improved compression ratios,
but also in improved performance in the downstream applications, such as
improved accuracy in sequence assembly that would be based on the lossy
rather than the original version of the quality values. This prospect and
its applications in high volume read sequencing is an exciting direction
for future investigations.

\section*{Acknowledgement}
\label{acknowledgement}
The authors would like to thank Golan Yona for helpful discussions
and Jill Banfield for devising the initial biological motivation for this work. This work
is supported by  Scott A.
and Geraldine D. Macomber Stanford Graduate Fellowship, Thomas and Sarah
Kailath Fellowship in Science and Engineering, 3Com Corporation Stanford Graduate
Fellowship, the La Caixa Fellowship, the EMBO long term Fellowship and
grants from the Center for Science of
Information (CSoI), and by an Innovation Research
Award from Hewlett Packard Labs.
\bibliographystyle{natbib}
\bibliography{jabref_genomics}
\end{document}